\documentclass[onecolumn,authoryear]{els-mrw} 

\usepackage{amsmath,amssymb,amsfonts,amsthm,makeidx,graphicx}
\usepackage{txfonts}
\usepackage{helvet}

\usepackage{caption}

\begin{document}

\chapter{History of the Observation of Stars}\label{chap1}

\author[1]{Andreas Schrimpf}%

\address[1]{\orgname{Philipps--Universität Marburg}, \orgdiv{Physics Department}, \orgaddress{Renthof 5, D--35032 Marburg, Germany}}


\maketitle
\begin{glossary}[Keywords]
star catalogs, accuracy of stellar positions, magnitudes of stars, distances to stars, double stars, astronomical photographic plates, time domain astronomy, variable stars, stellar spectroscopy
\end{glossary}

\begin{glossary}[Glossary]
\term{Armillary sphere} One of the oldest instruments to determine the position of celestial objects. A model of the sky consisting of the Earth in the center and rings for the equator and the ecliptic, as well as the celestial horizon.\\
\term{Heliometer} A double--image micrometer telescope. The two images can be moved with a micrometer to measure small angular distances with a very high precision.\\
\term{Meridian circle} An instrument for measuring the time of the passage of stars across the local meridian, i.e. the local sidereal time of stars. A telescope mounted on a fixed axis, so that it can be moved in the meridian plane only with a high-precision dial --- the meridian circle --- for determining the height of the observed object. \\
\term{Mural quadrant} A quadrant mounted onto a wall, oriented on the meridian. For measuring the position of a star, the time of a star transiting the meridian is observed, giving the right ascension. The declination is determined from the angular elevation of the star and the local geographic latitude.  \\
\term{Transit instrument} A smaller design than the meridian circle's.\\
\term{Objective prism} A prism mounted in front of the objective lens of a telescope, turning the telescope into a crude spectrometer for all stars in the field of view.\\
\term{Quadrant} An instrument to measure angles up to 90 degrees, used for measuring the altitude of a star or the angular distance between stars. 

\end{glossary}

\begin{glossary}[Nomenclature]
\begin{tabular}{@{}lp{34pc}@{}}
4MOST & 4-metre Multi-Object Spectroscopic Telescope\\
ALMA & Atacama Large Millimeter/submillimeter Array\\
CCD & Charge--Coupled Device \\
CMOS & Complementary metal-oxide-semiconductor\\
ESPRESSO & Echelle SPectrograph for Rocky Exoplanets and Stable Spectroscopic Observations\\
HARPS & High Accuracy Radial velocity Planet Searcher\\
HESS & High Energy Stereoscopic System\\
HIPPARCOS & High Precision Parallax Collecting Satellite\\
HCO & Harvard College Observatory\\
INTEGRAL & International Gamma-Ray Astrophysics Laboratory\\
LAMOST & Large Sky Area Multi-Object Fiber Spectroscopic Telescope\\
MAGIC & Major Atmospheric Gamma-Ray Imaging Cherenkov Telescopes\\
SDSS & Sloan Digital Sky Survey\\
\end{tabular}
\end{glossary}

\begin{glossary}[Key points]
\begin{itemize}
	\item The first \textbf{star catalog} was compiled by Hipparchos in the second century BC with an accuracy in stellar positions of about 0.5 degrees.
	\item The concept of \textbf{magnitudes} as a quantified measure of the brightness of stars was first mentioned by Manilius.
	\item New \textbf{mechanics} and \textbf{telescopes} helped to achieve an earthbound accuracy of less than 1 arc second for determining stellar positions.
	\item The first \textbf{stellar parallax} could be determined in 1839 by Friedrich Bessel.
	\item The dry \textbf{photographic plate} was introduced as the first detector in astronomy in the second half of the 19th century.
	\item \textbf{Stellar spectroscopy} began in the first half of the 19th century.
\end{itemize}
\end{glossary}

\begin{abstract}[Abstract]
	
There are about 6000 stars, that can be seen with the naked eye and have been observed for centuries for various purposes. More modern investigations using advanced telescopes show that our Milky Way, a quite common galaxy, consists of about 100 -- 400 billion stars. 
And, it is estimated that there are between 200 billion to 2 trillion galaxies in the observable universe --- all of them consist mostly of stars, and sending observable signals
which also represents nothing more than a superposition of the light of individual stars.
So we can conclude that the most common observable objects in the Universe are \emph{stars}.

In this chapter, we focus on the long history of the observation of stars (compared to studies in other fields of science) to find out more about the nature of these objects.

\begin{figure}[ht]
	\centering
	\includegraphics[width=.6\textwidth]{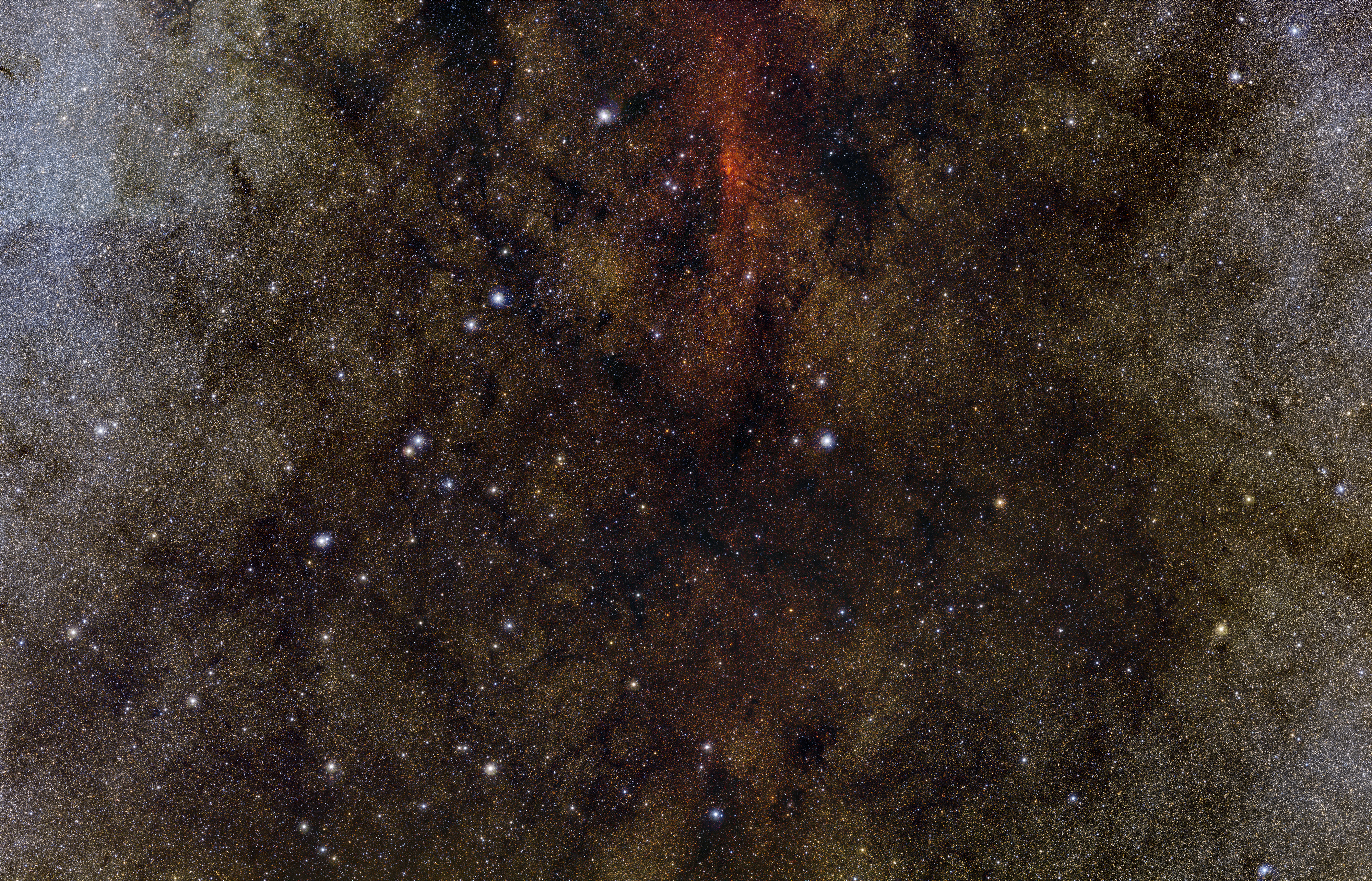}
	\caption*{One million stars — towards the dark heart of the Milky Way. (ESO/VISTA. Cambridge Astronomical Survey Unit, https://www.eso.org/public/images/eso0949b/)}
	\label{graphical_abstract:fig1}
\end{figure}

\end{abstract}

\section{Introduction}\label{sec_intro}
Observing the sky --- in the daytime and nighttime --- has been a part of human natural interest as long as mankind exists. The sequence of day and night as well as the sequence of seasons are obviously ruled by the relative position of the observer to the Sun. At nighttime the most prominent object is the Moon adding a monthly sequence of brighter and fainter nights due to its changing phases. More experienced observers will find other bright moving and changing objects, the planets of our Solar System. The nightly background to these objects is the arrangement of the stars, which were mistakenly called fixed stars by the Pythagorean philosophers.

For the Pythagorean philosophers and their successors the roughly 6000 fixed stars\footnote{The Bright Star Catalog \citep{Hoffleit1991} lists 9110 objects, however including objects below the human seeing limit as well as some supernovae.} formed a kind of background (map) of the moving and changing objects of the Solar System and thus their first interest in observing stars was to determine their \emph{position on a two-dimensional sphere}. Measurements of positions of stars with different or better methods increasing the accuracy of observations span the longest period discussed in this chapter and are still ongoing with the latest high precision machine, the Gaia space observatory \citep{Gaiamission}, which uses the \emph{parallax} to establish a three-dimensional map of stars in our galaxy.

With increasing the precision in determining the positions of stars, astronomers discovered that stars are not fixed in a certain position but show a \emph{proper motion}. The proper motion of the stars turned out to be a useful quantity for identifying and investigating structures of groups of stars. At higher resolution, telescopes revealed many of the known stars as \emph{double or multiple stellar systems}. This discovery is not only important to derive the \emph{mass} of the stars, but also for getting access to the \emph{evolution of stellar systems}.

In early times, the stars' \emph{brightness} was of interest only in easier describing asterisms or constellations. With the invention of the telescope, fainter stars were included in catalogs, increasing the need for more precise methods in comparing and finally in recording the brightness of stars. It was known since the Middle Ages that some stars show large variations in their brightness, but by continuously recording and comparing the brightness of many stars, thousands of \emph{variable stars} were discovered. By identifying different types of variable stars, a new field was introduced, the time domain astronomy.

Most of the physical properties of stars remained uncertain and hidden for a long time. 
Adopting spectroscopy to astronomical observations proved to be the key to accessing surface parameters such as temperature, chemical composition, density, and gravitational field strength. Applying these parameters to models, the \emph{age} of stars can be deduced. Together with 3D positions of stars on a galactic scale, this enables us to study the \emph{archaeology} of our galaxy.

In the sections of this chapter, we will follow the improvement in observations of stars on a \emph{chronological timeline} --- the \emph{History of the Observation of Stars} --- switching between improvements in the determination of different properties of stars.

\section{Ancient Observations}\label{sec_anc_obs}

For early cultures, celestial objects play an important role in helping to interpret good or bad events of daily life. The first astronomers are believed to be a kind of priest with a decent knowledge of celestial objects, especially of the motions of the Sun, Moon, and the planets of our Solar System. 
Stars are showing up in groups. The boundaries of these groups are not unambiguous, leading to different definitions of what we call constellations in different cultures. These constellations form a map of the sky, which is used to determine the position of the moving objects. The positions of stars were indicated by the description of the constellations.
All the ancient cultures in Mesopotamia, Greece, Egypt, India, China, or Mesoamerica used observations of the Sun and Moon to develop solar, lunisolar, or lunar calendars and thus relied on more or less precise positions of Solar system bodies relative to stars. Therefore, traces of such stellar maps can be found in many ancient cultures, for reference see \cite{ruggles2015}.

\subsection{The first star catalog}\label{sec_anc_obs:first_cat}
It was Hipparchos (c.\ 190 BC -- c.\ 120 BC), who in the 2nd century BC compiled the first \emph{star catalog} we are aware of. Hipparchos is ranked among the greatest ancient astronomers, but little is known about his life \citep{linton2004}. Only one of his publications and essays still exists, that is a commentary on the \emph{Phenomena of Eudoxus and Aratus}. Our knowledge of Hipparchos' work relies on second-hand reports, the most famous one is Ptolemaios' \emph{Almagest}; another source is \emph{Naturalis Historia} by the Roman philosopher Pliny the Elder.  And, it is well accepted that Hipparchos' work --- the work of the antique Greek astronomers --- was influenced by Babylonian astronomy \citep{Hoffmann2017}.

\begin{figure}[ht]
	\centering
	\includegraphics[width=.45\textwidth]{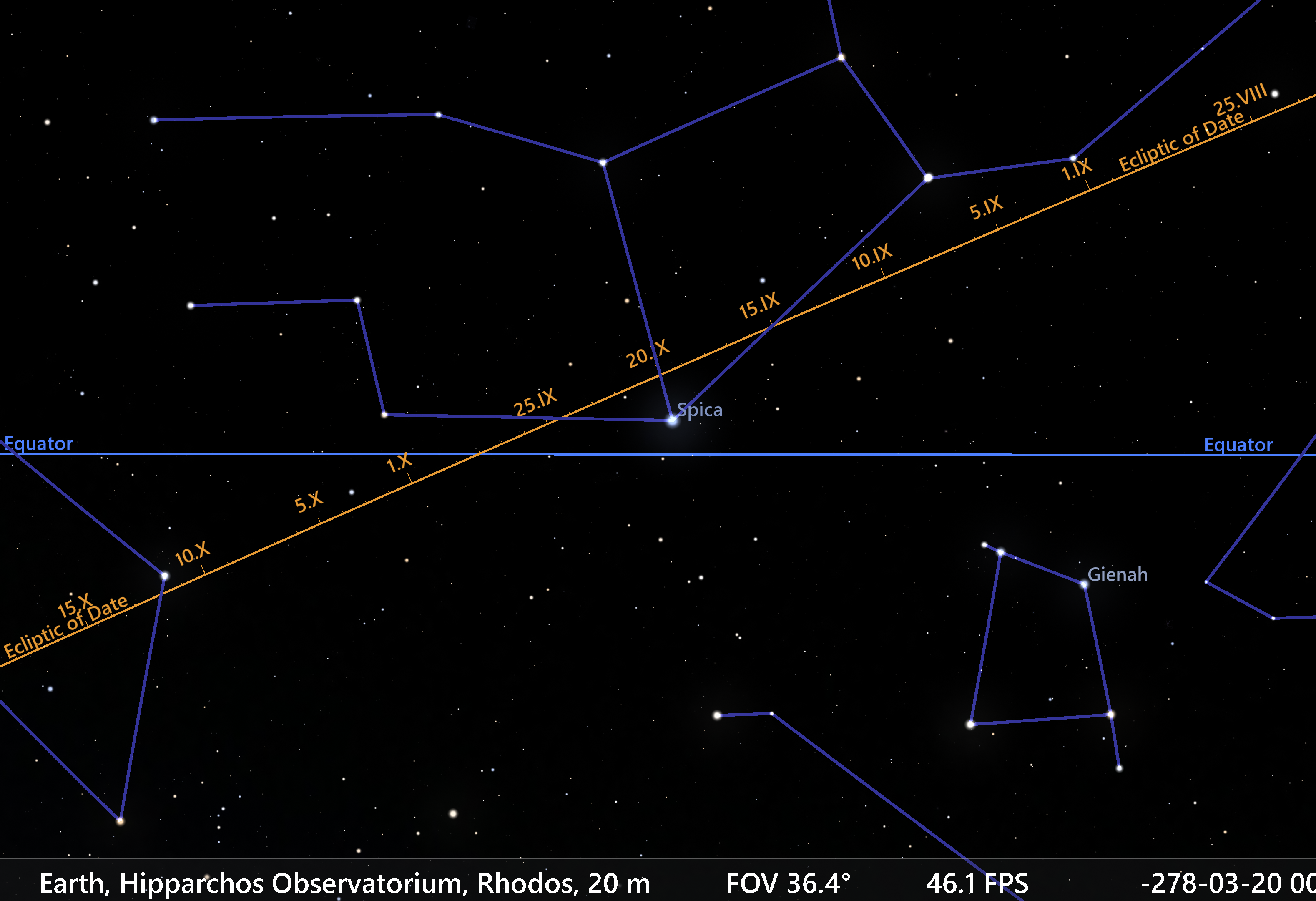}\hspace{1mm}
	\includegraphics[width=.45\textwidth]{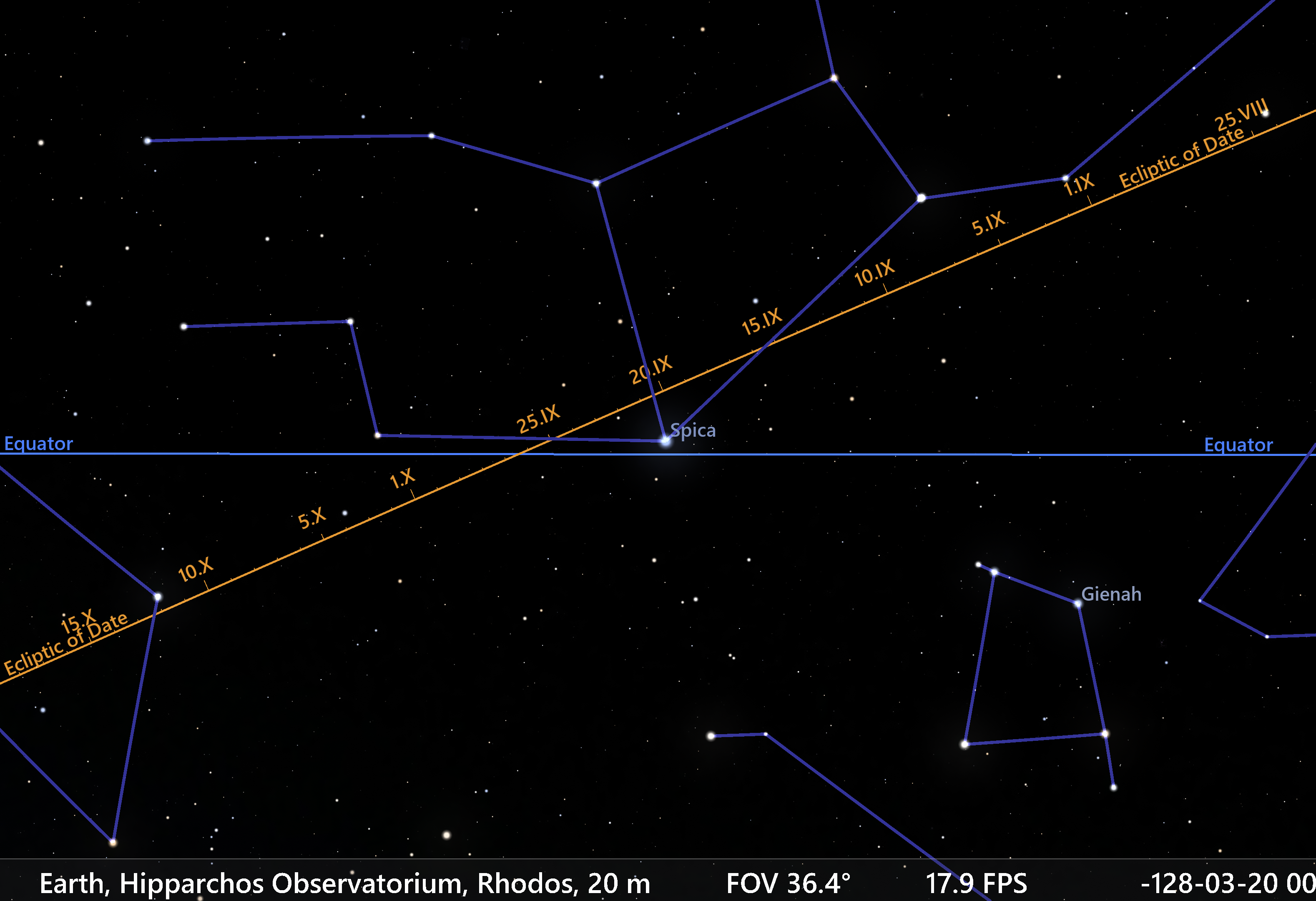}	
	\caption{Autumn equinox as seen from Rhodes 150 years before Hipparchos (left) and at the time of Hipparchos' observations (right). The ecliptical length of Spica differs by 2.03 degrees. (Figures prepared with Stellarium.)}
	\label{sec_anc_obs:precession}
\end{figure}

Hipparchos used an observatory at Rhodes, Greece, with instruments like armillary spheres, sundials, and water clocks. About 127 BC he completed the compilation of a catalog of 850 stars with positions given in ecliptical coordinates, which covers the part of the sky visible from his observatory. Hipparchos catalog is lost, but a reconstruction could be created from the descriptions of the constellations in his commentary on the \emph{Phenomena} by \cite{Hoffmann2017}. It is worth noting that recently a small part of Hipparchos' catalog was discovered hidden in a medieval parchment manuscript \citep{Gysembergh2022}. We have no information about the accuracy of Hipparchos' observations. But, he is credited with discovering the precession of the equinoxes by comparing the ecliptical lengths of Spica and a few further stars of his own measurements with observations 150 years earlier. The effect of the precession during that time span is roughly 2 degrees (Fig. \ref{sec_anc_obs:precession}), so the accuracy of Hipparchos' observations must have been better than that.

The oldest star catalog that still exists was published by Ptolemaios (c.\ 100 -- c.\ 170 AD) in his \emph{Almagest}. In this fundamental work about the knowledge of astronomy at that time, Ptolemaios gave much credit to Hipparchos. He compiled his catalog of 1022 stars for the epoch of 137 AD. It is generally accepted among historians that Hipparchos' catalog was the base of Ptolemaios' compilation, supplemented by further stars and supported by his own observations, presumably from Alexandria, Egypt.
Ptolemaios described his instrument and methods of observations and estimated the measurement errors to about $1.5^\circ$ to $2^\circ$ \citep{Hoffmann2017}. A comparison of this catalog with the modern HIPPARCOS catalog reveals an error of less than $0.5^\circ$ in both, ecliptical longitudes and latitudes \citep{verbunt2012}.

\subsection{The brightness of stars in ancient catalogs}\label{sec_anc_obs:anc_mag}
 
\begin{figure}[ht]
	\centering
	\includegraphics[width=.9\textwidth]{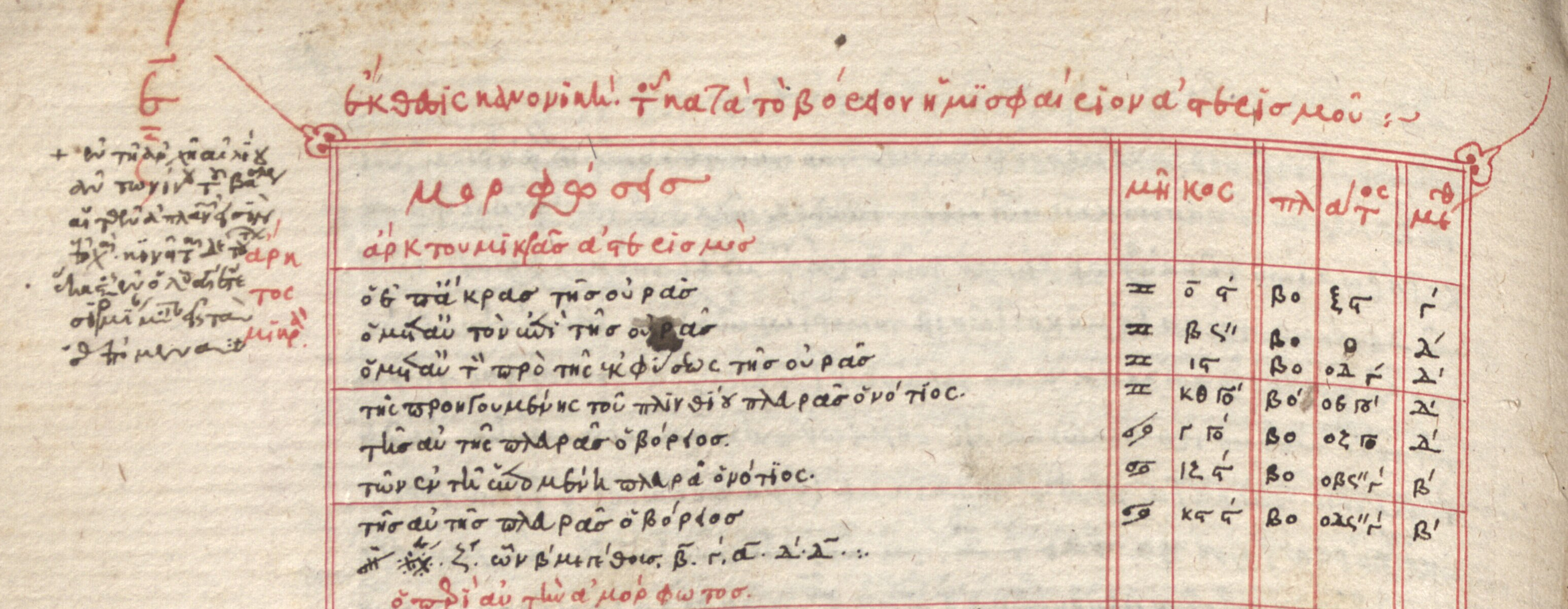}
	\caption{Stars of the constellation Ursa Minor (UMi) from \emph{Almagest}, page 324. The left column contains a description of the stars, and the data to the right are longitude, latitude, and magnitude (Digitized by Digitale Bibliothek, Münchener DigitalisierungsZentrum).}
	\label{sec_anc_obs:umi_almagest}
\end{figure}

Hipparchos' data of the constellations listed in the above-mentioned commentary on the \emph{Phenomena} only have rudimentary information about the brightness of stars. The concept of \emph{magnitudes}, a quantified measure of the brightness of stars, is confirmed for the first time in the poem \emph{Astronomica}, written by Manilius, a Roman poet of the 1st century AD. Ptolemaios most probably continued to use this system \citep{cunningham2020}. In his catalog, he includes mostly numerical data for the brightness of stars, the brightest one starting with 1 mag, and the faintest stars visible with the naked eye with 6 mag, as well as a few descriptive data. As an example, Fig. \ref{sec_anc_obs:umi_almagest} shows the page from \emph{Almagest} listing the stars of the constellation Ursa Minor (UMi). Neither a method nor an instrument to determine the stellar magnitudes is mentioned in the \emph{Almagest} or any other literature of that time. Most likely, they used some kind of naked-eye estimation.

The modern magnitude scale was defined by Norman Pogson in 1856 (section \ref{sec_telescope_era:more_properties:brightness}) so that it matches the use of ancient astronomers. A comparison of modern data with the listed brightness of stars in Ptolemaios' catalog results in an accuracy estimation of 1.3 to 1.6 mag \citep{protte2020}.

\section{The Beginning of the Modern Era}\label{sec_beg_mod_era}

\subsection{The catalogs in the following centuries}\label{sec_beg_mod_era:cat_old}

The usage of ecliptical coordinates in Ptolemaios' catalog enabled an easy transformation to epochs of later times by just adding the precession constant times the number of years that passed by to the longitude of the stars. So for a long time, no efforts were made to take new measurements of coordinates. In 964 AD the famous astronomer Al-Sufi (903 -- 986) in his \emph{Book of Fixed Stars} compiled a catalog of identical stars as in Ptolemaios' list by converting them to a new epoch and assigning magnitudes from his own observations at Isfahan \citep{protte2020}.

\begin{figure}[ht]
	\centering
	\includegraphics[width=.6\textwidth]{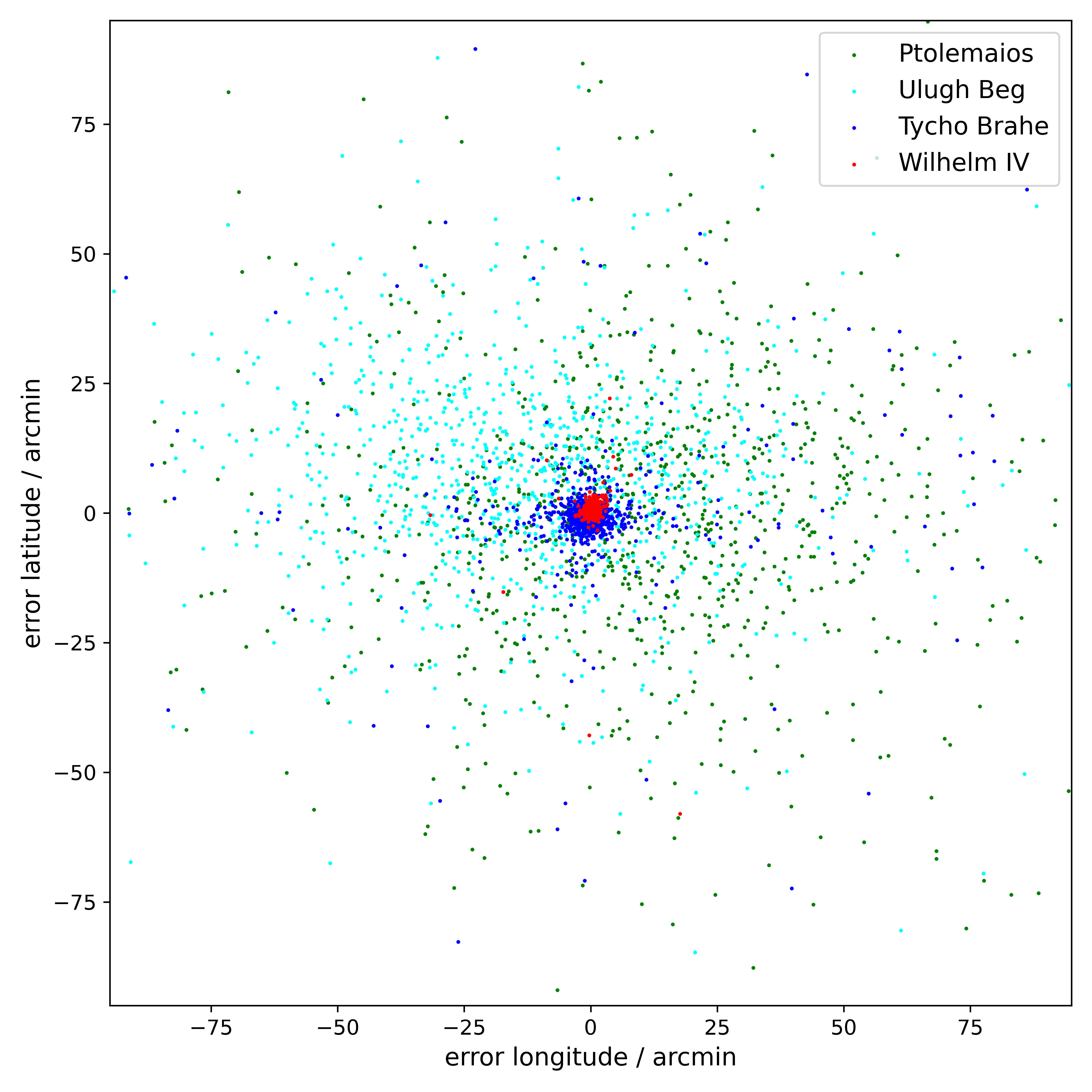}
	\caption{Comparison of the accuracy of star catalogs of the pre--telescope times. Shown are the differences in coordinates from the pre--telescope catalogs and calculated coordinates from the HIPPARCOS catalog. The accuracy of Wilhelm's catalog surpasses that of all other pre--telescope catalogs.}
	\label{sec_beg_mod_era:accuracy}
\end{figure}

Over the centuries, however, the errors in the copies of Ptolemaios' catalog accumulated. In the first half of the 15th century, Ulugh Beg (1394 -- 1449) redetermined the position of stars from Ptolemaois' and Al-Sufi's catalogs, which were accessible from his new observatory at Samarkand. He made efforts to increase the accuracy, especially of the declination measurements, by increasing the size of the instrument: he constructed the well-known Fakhri sextant with a radius of about 36 m and thus was able to improve the accuracy of the catalog roughly by 30\% (Fig.\ \ref{sec_beg_mod_era:accuracy}).

\subsection{Magic mechanics improving the accuracy}\label{sec_beg_mod_era:mechanics}

The idea of Ulugh Beg to create better instruments did not lead to a breakthrough. Unfortunately, Ulugh Beg's observatory was almost entirely destroyed within a few generations after his death, we have no good knowledge about the power and the problems of his instruments.

But by the end of the 16th century, two major improvements in mechanics led to instruments that really changed the game. The credit for compiling the most accurate catalog in pre--telescope times goes to Wilhelm IV (1532 -- 1592), the Landgraf von Kassel, and his coworkers \citep{verbunt2021}. Wilhelm IV founded a modern observatory at Kassel in 1560 to improve the positional data of stars, especially for calendar purposes. He hired the mathematician and astronomer Christoph Rothmann (1550/60 -- 1601) and the brilliant instrument maker Jost Bürgi (1552 -- 1632) \citep{schrimpf2021}. One of the many achievements of Bürgi is a major improvement in the accuracy and synchronization of clocks. He is known as the \emph{inventor of the seconds}, as his clocks were the first ones with a subsecond accuracy! Wilhelm's coworkers determined most of the right ascensions via \emph{time measurements} of stellar positions, i.e., the sidereal time, in the meridian and outside the meridian using these clocks and in 1586 compiled a catalog of 383 stars. They achieved an accuracy of about 1 arcmin, more than an order of magnitude better than former catalogs and \emph{two times better than that of Tycho Brahe's catalog}. The Wilhelm IV catalog is the first one that lists positional data in equatorial (and ecliptical) coordinates. However, it first appeared in print, edited by Curtz in 1666, long after the dissemination of Tycho Brahe’s catalog, and as a result, it had a rather limited impact.

Tycho Brahe (1546 -- 1601) is known for building high-precision instruments. 
In his early years, Tycho Brahe traveled through Europe, taking studies at many universities and starting astronomical observations. Inadequate observation methods of observatories at the time led to his early involvement with the methodology and instruments for the precise measurement of celestial positions. 
In April 1575 Tycho Brahe visited the already experienced observer Wilhelm IV and studied his instruments. In the correspondence that followed, Wilhelm IV in turn profited from suggestions for improved measurement accuracy by Brahe, e.g., transversal lines for better reading accuracy. Instead of using clocks\footnote{Tycho Brahe did not believe that clocks could have adequate accuracy. He was proved wrong by Jost Bürgi.} Brahe designed larger instruments than Wilhelm IV mostly made of brass with high precision scales; the most accurate instrument --- mechanically well below one arcmin --- was a mural quadrant. Brahe compiled a catalog of 777 stars, which was published in 1602, one year after his death. The full catalog of 1004 stars was published by Kepler in 1627 \citep{verbunt2010}. This catalog has an accuracy slightly larger than 2 arcmin.

\section{Beginning of the Telescope Era}\label{sec_telescope_era}
\subsection{Telescopes used for positional astronomy}\label{sec_telescope_era:pos_astronomy}

The first usage of telescopes in astronomy was introduced by Galileo Galilei and Thomas Harriot in 1609. In the first half of the 17th century, astronomers concentrated on observing solar system bodies and sunspots, not stars. 

One of the early star catalogs from observations using a telescope was compiled by Johannes Hevelius (1611 -- 1687), and printed by his wife after his death in 1690. The accuracy of Hevelius' position measurements is similar to that of Brahe \citep{verbunt2010hevelius}.

John Flamsteed (1646 -- 1719), the first \emph{Astronomer Royal} and founder of the Royal Observatory at Greenwich, UK, mounted a telescope on a sextant and later on a mural quadrant, thus improving the accuracy of angular distances between stars and absolute positions determined with the mural quadrant and a clock. His catalog of 2934 stars was published posthumously in 1725 and had an accuracy of better than 40 arcsec \citep{lequex2014}. Other catalogs followed throughout the 18th century using the same type of instruments from La Caille (1713 -- 1762) and Mayer (1723--1762) with some progress in accuracy \citep{lequex2014}.

\begin{figure}[ht]
	\centering
	\includegraphics[width=.6\textwidth]{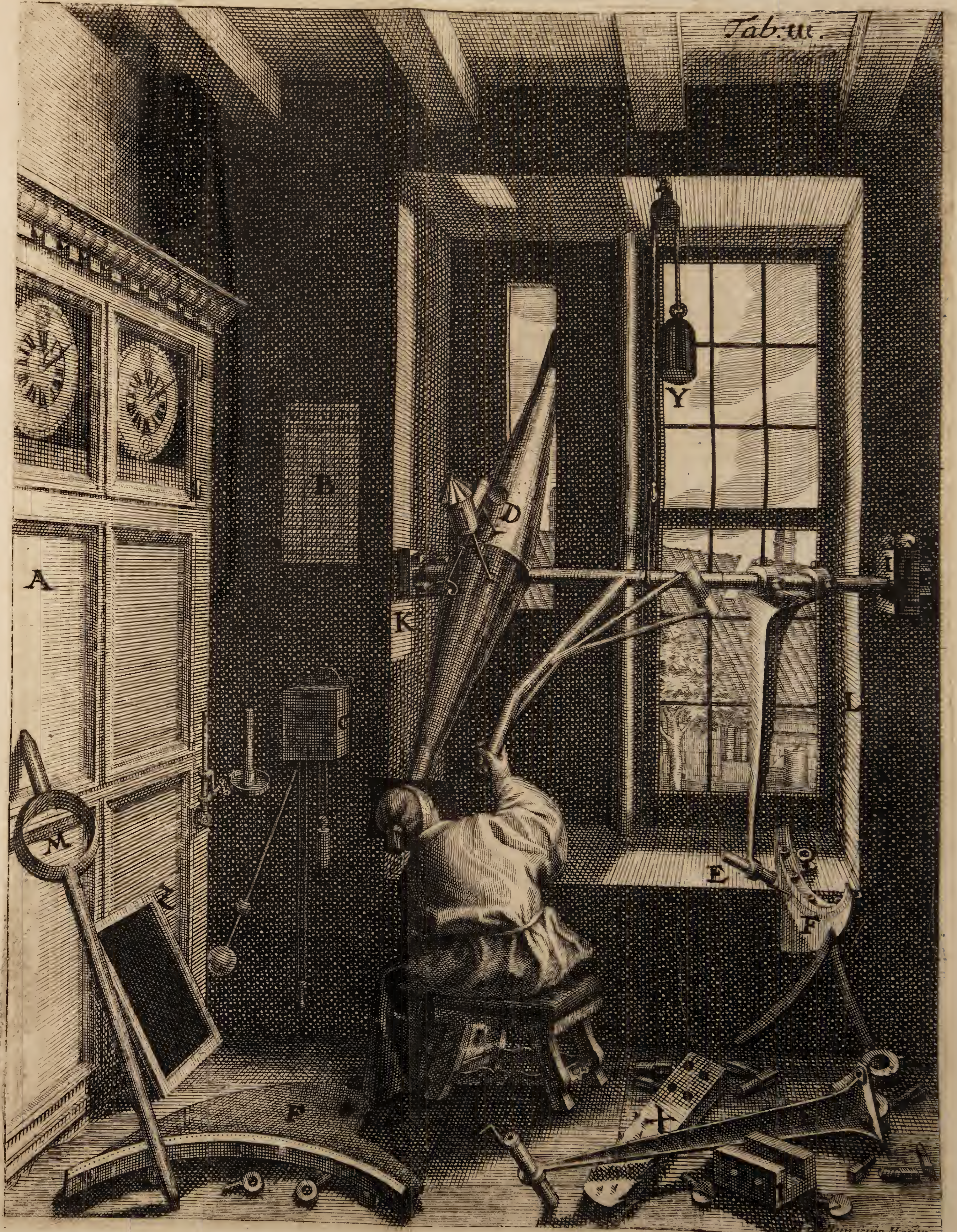}
	\caption{Olaf R{\o}mer invented the meridian circle in 1690 and achieved an accuracy of 4 arcsec for stellar positions \citep{horrebow1735}}.
	\label{sec_telescope_era:meridian_circle}
\end{figure}

It was Olaf R{\o}mer (1644 -- 1710) who realized the power of an optical instrument to further improve the accuracy of stellar position measurements. He mounted a telescope on a fixed axis so that it could be moved in the meridian plane only with a high precision dial for the height of the telescope's direction --- the first \emph{meridian circle} (Fig.\ \ref{sec_telescope_era:meridian_circle}). The key differences to the mural quadrant are the larger telescopes, that can be used with meridian circles and the sturdier mount of the moving axis. Unfortunately, most of R{\o}mer's documents of his astronomical work were destroyed in a fire except for observations of 88 stars, which were analyzed about 150 years later by Galle. With his instrument  R{\o}mer achieved an accuracy of 4 arcsec \citep{lequex2014}.

In the 19th century, Friedrich Wilhelm Argelander (1799 - 1875) was able to further enhance the positional accuracy to 0.9 arcsec \citep{eichhorn1974} using a meridian circle and carefully correcting precession, nutation, and aberration of the apparent position of stars. The \emph{Bonner Durchmusterung}, a catalog of 342,198 stars he compiled from his observations, was published between 1859 and 1862.

\subsection{High precision measurements of stellar positions -- fundamental star catalogs}\label{sec_telescope_era:fundamental_stars}

In the 18th century, astronomers realized that it was necessary to determine the absolute position of a few stars with the highest precision and use these as references for filling in thousands of stars from differential observations, as Wilhelm IV and Tycho Brahe had done earlier. Such independently observed stars are called \emph{fundamental stars}.

Nevil Maskelyne (1732 -- 1811), the fifth British Astronomer Royal, picked 36 bright zodiacal stars from Bradley's observations, which were observed in the daytime with the Sun, and for several years continuously determined the right ascension using a meridian circle \citep{knobel1877}. He improved the time measurements of the transit of stars to about 0.1 seconds.

Friedrich Bessel (1784 -- 1846) identified offsets in absolute positional measurements due to different reaction times of observers, a systematic error caused by observations by eye, the so-called personal equation, and after carefully reducing his observations of the same 36 stars, that Maskelyne, Piazzi and others had observed, he published \emph{the first catalog of fundamental stars, the Tabulae Regiomontanae} in 1830 \citep{bessel1830} with an accuracy of 0.7 arcsec \citep{fricke1985}.

In the second half of the 19th century, Arthur von Auwers (1838 -- 1915) presented the \emph{Fundamental Catalog -- FC}, the first of the so-called ''German Series'' of fundamental catalogs in order to define a \emph{reference system} for positions of astronomical objects. The final catalog in the series was the FK5 catalog, a compilation of about 260 individual catalogs, observed mostly with meridian circles and some astrolabes \citep{perryman2012}. The catalog was published in 1988. It contained 1535 stars and established an accuracy of 0.04 arcsec \citep{hoeg2017}.

Further improvement in the accuracy of positional astrometry could be achieved by using space telescopes (section \ref{sec_modern_astronomy}).

\subsection{Further properties of stars emerge}\label{sec_telescope_era:more_properties}

\subsubsection{The discovery of the proper motion}

As we know from a discussion by Ptolemaois in his \emph{Alamgest}, it was Hipparchos who thought about a relative motion of the presumably fixed stars. He argued that proof was required to claim that the stars are fixed on a sphere. Hipparchos compared the relative positions of stars in constellations and alignments with observational data 150 years earlier and could not find any hint of a relative change in the positions of stars \citep{verbunt2019}. Ptolemaois could not find any hint either another 150 years later. However, the problem remains open throughout the centuries.

In 1717, Halley compared observations of four stars by Hipparchos and Ptolemaios with those from Brahe and others and despite a warning from Brahe, that the data might be incorrect, concluded that all four stars had shown proper motion. But he was wrong.

It was Jacques Cassini (1677 -- 1756), known as Cassini II, who showed that the ancient observations were too inaccurate to be used for calculating the proper motion. Cassini corrected Brahe's observations and compared them with his own data of some stars and thus provided the first significant evidence for the proper motion of a star -- in this case of Arcturus \citep{verbunt2019}.

\subsubsection{The discovery of double stars}

There are some records of observations of close stars in pre--telescope era, but it was the use of telescopes that led to deeper insights into the field of stellar systems. Castelli in 1617 discovered that Mizar is a double star, in 1664 Hooke identified $\gamma$ Ari as a double star. Within the first century of telescope observations 10 such close pairs of stars have been detected, separated by a few arc seconds \citep{aitken1918}. 

The open question after the discovery of the laws of gravity by Isaac Newton in the 17th century whether pairs of stars are \emph{optical} or \emph{physical} pairs was first discussed by John Michell (1725 --1793). He calculated the probability of finding two stars very close to each other in the sky, presuming a random spatial distribution and given the total numbers of stars of any particular apparent magnitude \citep{michell1767}. He showed that this probability is rather small and thus concluded that most of the observed pairs of stars are not by chance in coincidence in direction, but are \emph{real physical pairs bound by gravity}.

The first catalogs of double stars were published by Christian Mayer (1719 -- 1783) in 1781 and Wilhelm Herschel (1738 -- 1822) in 1782.
In 1803 Herschel demonstrated from his own observations that certain double stars are true binary systems: the relative changes in positions could only be explained by an orbital motion.

\begin{figure}[ht]
	\centering
	\includegraphics[width=.6\textwidth]{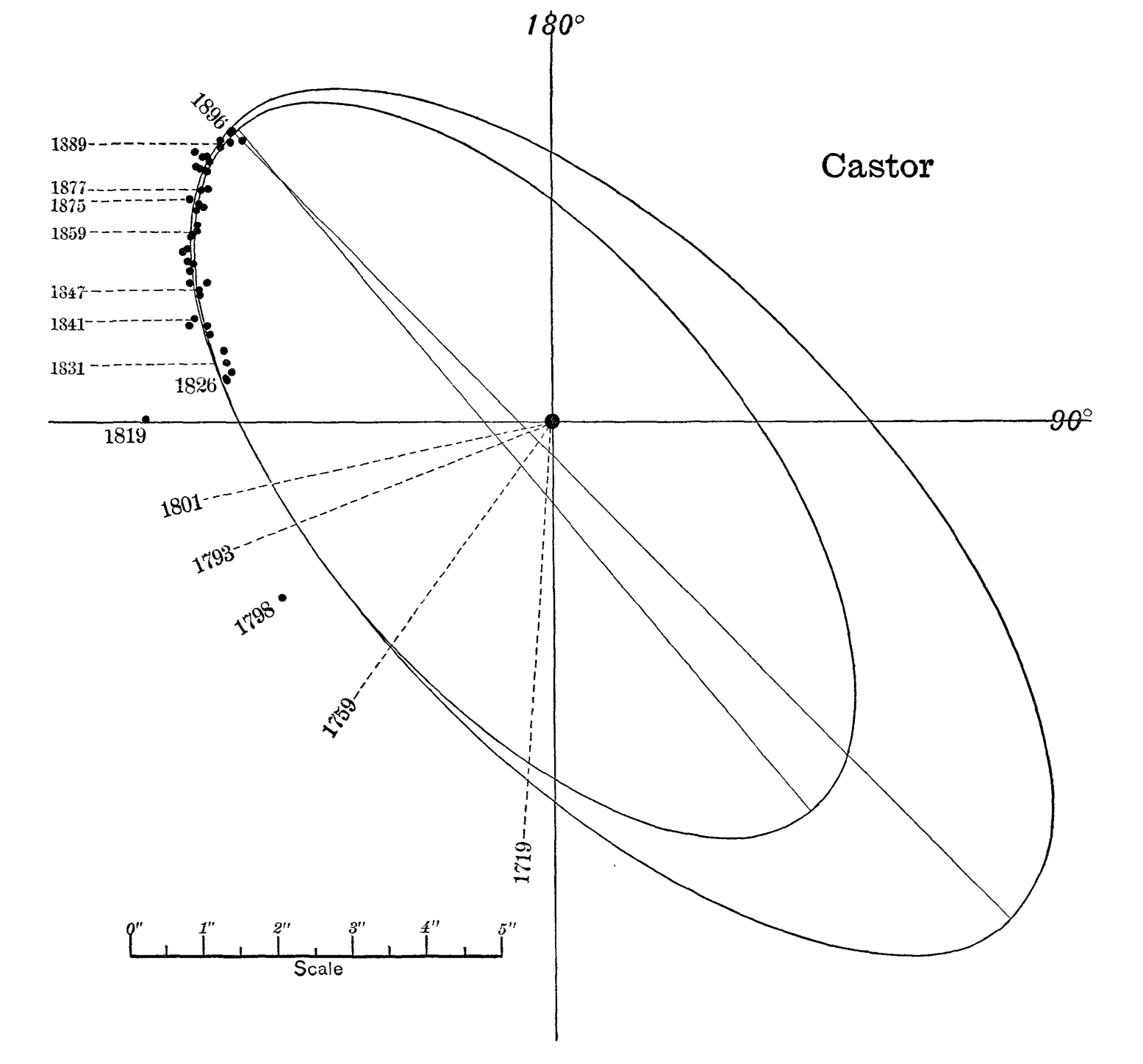}
	\caption{Relative orbit of the double star system Castor A and B, observations from  1826 to 1889. (Popular Astronomy 4, 1896, p. 286)}
	\label{sec_telescope_era:orbit_castor}
\end{figure}

The visual binaries, i.e. gravitationally bound binary systems, that can be visually resolved into two stars, perform Kepler orbits around their center of mass (Fig.\ \ref{sec_telescope_era:orbit_castor}). Thus, the relative \emph{masses} of the two stars can be determined from the orbits. To calculate the absolute mass, however, the distance to the stellar system is required.

\subsubsection{The brightness of stars}\label{sec_telescope_era:more_properties:brightness}

Determining the brightness of stars remained an unsolved problem until detectors with reproducible sensitivity became available at the end of the 19th century. However, Ptolemaios' magnitude scale was well established for the range of naked-eye vision. It turned out to be a scale of the ratio of intensities, a logarithmic scale. However, the scaling factor was not fixed yet, and not agreed upon between astronomers.
With telescope observations, the magnitude scale had to be extended to fainter stars. The typical procedure was to reduce the aperture of the telescope to a size, where a certain star would just not be visible anymore. That aperture size was used as a measure of the magnitude.

In 1856 Norman Pogson (1829 -- 1891) reviewed the magnitude scales of various astronomers of his time and in the end came up with a definition of the scale factor that still holds: 5 magnitudes correspond to a brightness ratio of 100 \citep{pogson1856}. 
Still, the absolute calibration of the magnitudes of stars was an open question at that time. This was 
solved with the invention of detectors for the brightness of stars.

With ongoing observations of more and more astronomers, the \emph{variable stars} came into their focus. Some stars change their brightness on different time scales. Historical records of this phenomenon were on Algol by ancient Egyptian astronomers and on Mira by Fabricius in the 16th century. For determining the magnitude of variable stars, Argelander in 1844 proposed a very precise step estimation method \citep{argelaner1844} (comparing a star of unknown brightness with several neighboring known stars), which is still used today for naked-eye observations. Special telescopes were designed that allowed easy comparison of two stars, e.g. the Harvard Meridian Photometer.

\subsubsection{The distance to the stars} \label{sec_telescope_era:more_properties:parallax}

The distances to astronomical objects have puzzled astronomers for centuries. The reason: they are incredibly larger than any distance on Earth that humans experience in their daily lives. The lunar horizontal parallax is about 1 degree and was known to ancient scientists. But the parallax to any other astronomical object could not be determined before telescopes became available.

At the beginning of the 19th century, several observatories tried to find parallaxes of stars by very precise positional measurements throughout the year using the radius of the Earth's orbit as a baseline. However, the accuracy of meridian circles was not sufficient to observe stellar parallaxes. 

With the increasing interest in studies of double star systems, the desire for precise measuring devices for very small angular distances arose. The answer was the invention of the \emph{heliometer}, a double-image micrometer telescope, in the 18th century. It was improved by Fraunhofer in the 1820s. Bessel was impressed by the power of Struve's heliometer at the Dorpat observatory. In 1829, he put a new heliometer into operation at his observatory in Königsberg and observed the position of the double star system 61 Cyg in 1837 and 1838 relative to two other stars with zero proper motion or parallax. Bessel chose 61 Cyg because it showed a large proper motion which suggested a small distance and thus a large parallax. He was able to find the typical helical motion of 61 Cyg and determined a parallax of about 0.3 arc seconds \citep{bessel1839}, the first parallax observation of a star. It's worth mentioning that Thomas Henderson (1798 -- 1844) and Friedrich Georg Otto von Struve (1793 -- 1964) were pretty much on a par with Bessel in this ''space race'' in the 1830s.

\section{Detectors for Recording Data}\label{sec_detectors}

Before recording devices became available any observation was affected by the skills of the observer. Additionally, errors in the detection could only be determined by carefully comparing them with simultaneous observations of others. A huge breakthrough was the invention of photography in the second half of the 19th century. 

\begin{figure}[ht]
	\centering
	\includegraphics[width=.6\textwidth]{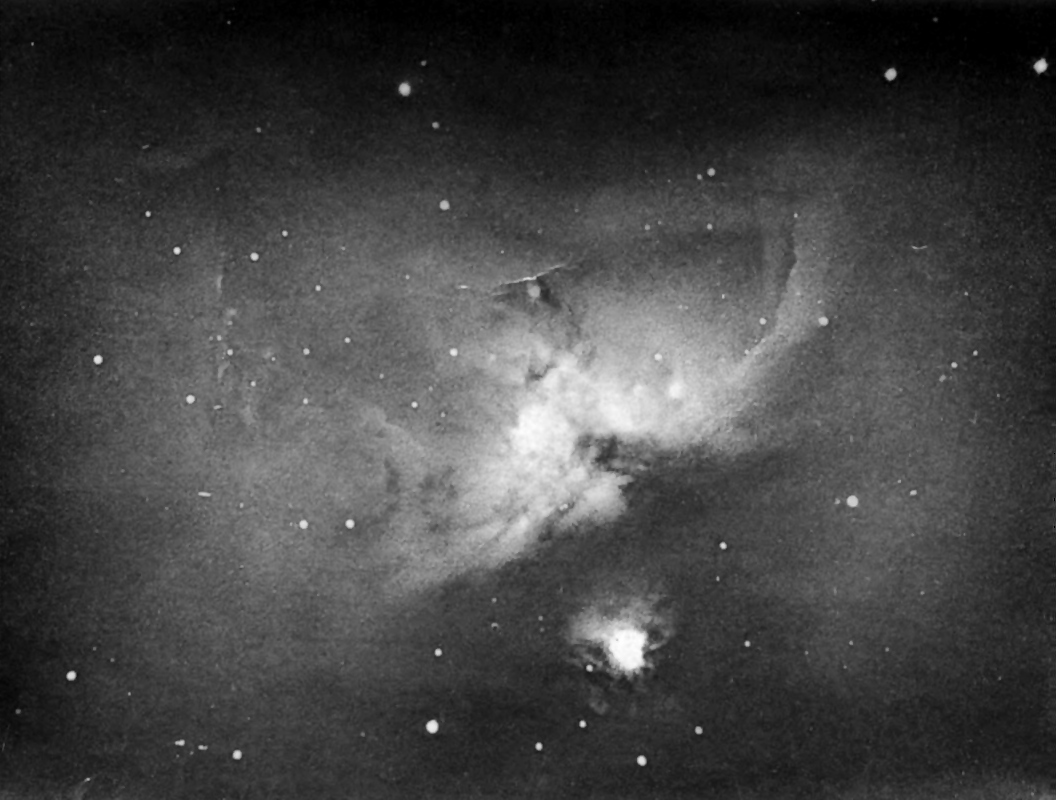}
	\caption{Scan of the 1883 photo of the Orion Nebula made by Andrew Ainslie Common (1841–1903) for which he won the Gold Medal of the Royal Astronomical Society in 1884. Common's images (exposure time up to 60 min)  for the first time showed stars too faint to be seen by the human eye (Wikipedia).}
	\label{sec_detectors:astrophotography}
\end{figure}

After a few different developments, the glass plate with a dry light-sensitive emulsion prevailed. Although the quantum efficiency of the emulsions was in the range of a few percent, only, by integrating the number of incident photons with typical exposure times of 20 to 40 minutes or longer, stars and other objects too faint to be seen by the human eye become visible (Fig. \ref{sec_detectors:astrophotography}).

The professional use of photographic plates for scientific applications in astronomy was introduced at Harvard College Observatory by Edward Charles Pickering (1846 -- 1919) in 1882. Max Wolf (1863 -- 1932) began to exploit the new technique at Heidelberg Observatory in 1891. Many observatories were soon convinced of the great benefits of astronomical photographic plates as detectors and data storage devices.

In 1887 Ernest Mouchez (1821 --1892), the director of Paris Observatory, initiated the \emph{Carte du Ciel} and the \emph{Astrographic Catalogue} (AC),  a campaign to catalog and map the positions of millions of stars as faint as 11th or 12th magnitude by use of photographic plates. 20 observatories around the world participated and took about 22,000 plate images of the sky between 1895 and 1920. However, despite many efforts, the project never was completed. The first version of the catalog was released in 1997, and the second one in 2001. The positions of stars in the AC helped in combination with modern epoch positions to determine accurate proper motions. Two further surveys using photographic plates are worth mentioning, the first and second Palomar Observatory Sky Survey, POSS and POSS-II. Both surveys have been used to create catalogs of many different object classes.

\subsection{Time-domain astronomy}\label{sec_detectors:time_domain_astronomy}

With the usage of photoplates to store information about astronomical objects a new field in research was opened, the \emph{time domain astronomy}. With storing data taken at different dates any changes of objects --- in position and brightness --- could be continuously monitored and studied. The Harvard College Observatory is the first to start continuous observation campaigns in high spatial resolution (approx.\ 1-2 arc seconds) and low resolution but a large field of view (approx 15 arc seconds and roughly 20 degrees $\times$ 20 degrees) \citep{hco1971}. In Germany, the largest similar campaigns in terms of number of exposures were started by Cuno Hoffmeister (1892 -- 1968) in 1925 at Sonneberg Observatory \citep{kroll2009}. These campaigns were continued at both observatories until in the 1990s the CCD--detectors replaced the photoplates (see section \ref{sec_modern_astronomy}).

Time-domain astronomy studies in stellar astrophysics mainly investigate many kinds of variable stars. One of the most famous examples is the discovery of the period-luminosity relation for Cepheids by Henrietta Swan Leavitt (1868 -- 1921). During their campaigns both observatories, Harvard and Sonneberg, discovered and classified more than 10.000 variable stars, each.

The most important legacy of photographic plates is the long time span since the beginning of these campaigns --- more than 140 years up to now, which helps study the long-term properties of stars and search for optical transients. The rather long exposure time needed due to the low quantum efficiency is limiting studies of rapid changes, e.g. flare stars and optical transients of gamma-ray bursts.

\subsection{Stellar spectroscopy}\label{sec_detectors:stellar_spectroscopy}

Stars show up in different colors. The colors of bright stars were known to ancient astronomers and Aristole's view was, that light (of different colors) was something, all bodies could have.  In the 11th century, Ibn al-Haytham studied the refraction and dispersion of (solar) light using pieces of glass, the dawning of optical spectroscopy. At the beginning of the modern era prisms were known as dispersive optical components and were used by several famous scientists, e.g. Newton, Herschel, and others, to explore the nature of light. As the Sun is the natural source of light some of the main features of the spectrum of the Sun were known till the end of the 18th century.

William Hyde Wollaston (1766 -- 1828) was the first to discover the dark lines in the solar spectrum in 1802. These lines independently were rediscovered in 1814 and studied in more detail by Joseph von Fraunhofer (1787 -- 1826) using a prism spectrometer. Kirchhoff and Bunsen in 1859 associated these lines with absorption lines of atoms in upper layers of the solar photosphere --- the beginning of the chemical analysis of the outer part of stars! Almost simultaneously in 1862, William Huggins (1824–1910) in London, Father Angelo Secchi (1818 - 1878) in Rome, and Lewis M. Rutherfurd in New York started to analyze spectra of stars and nebulae.
Huggins used a slit spectrometer at the focal point of his telescope. Secchi started with slit spectrometers but finally chose an objective prism, a prism mounted in front of the objective lens of the telescope, turning the telescope into a crude spectrometer for all the stars in the field of view. Secchi classified about 4000 stars into 5 classes differing in strength and appearance of lines of different elements and molecules \citep{clerke1902}. It is worth noting that Huggins employed photography, but Secchi did all of his work visually and made the drawings of stellar spectra manually!

\begin{figure}[ht]
	\centering
	\includegraphics[width=.45\textwidth]{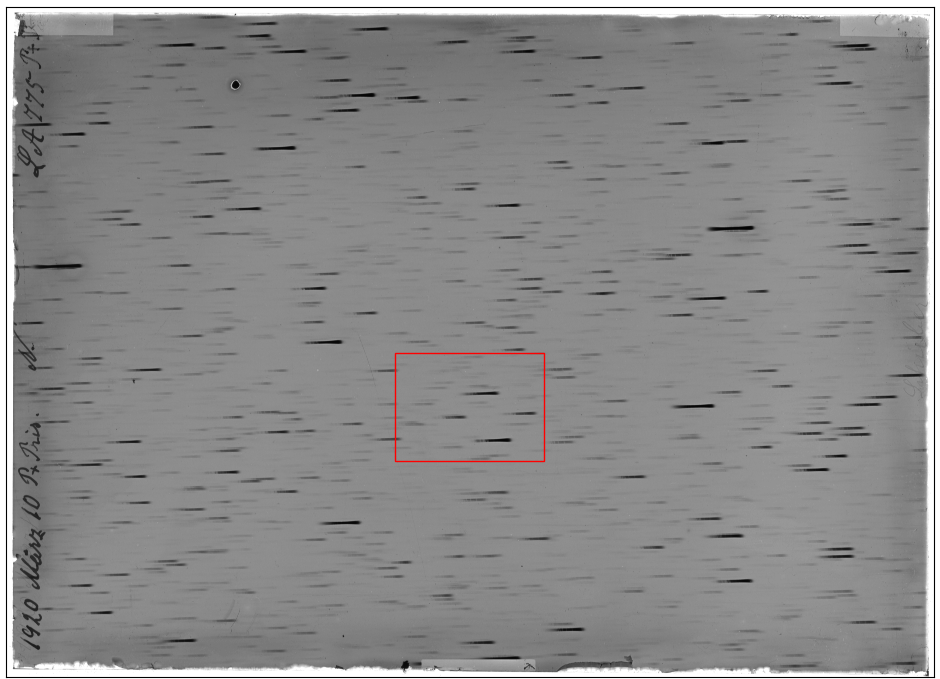}\hspace{1mm}
	\includegraphics[width=.45\textwidth]{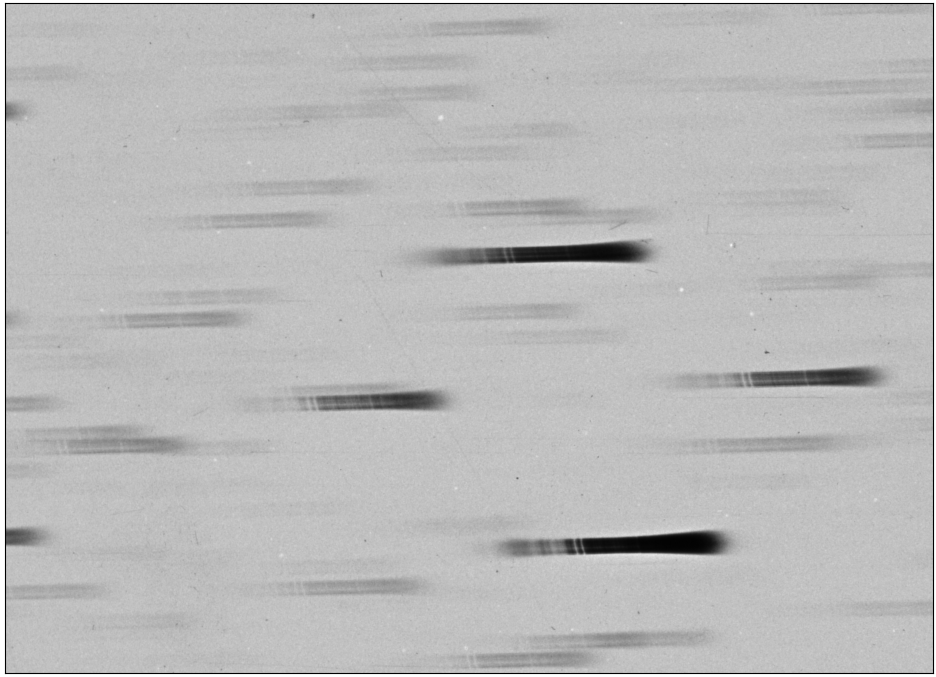}	
	
	\caption{Photoplate LA775, Hamburg Observatory, recorded with a 2-deg objective prism at the Lippert--Astrograph, 10. March 1920. The 1.5 m telescope was pointing to W UMa, the exposure time was 4 hours. W UMa, at that time, was a famous variable star --- a low-mass contact binary. In the plot on the left, the red rectangle marks a subsection, which is shown in detail on the right. This plate is accessible via the German plate archive, APPLAUSE \citep{Enke2024}.}
	\label{sec_detectors:spectral_photoplate}
\end{figure}

At Harvard College Observatory Pickering's main focus was on spectroscopy of stars! 
In memory of Henry Draper in 1886 Mrs. Draper made a generous donation to finance the spectroscopic investigation at the HCO named the \emph{Henry Draper Memorial} \citep{hco1971}. Applying an objective prism to one of the telescopes, on January 26, 1886, the first spectroscopic exposure of a plate recorded the spectra of about 40 of the Pleiades simultaneously! As an example, Fig. \ref{sec_detectors:spectral_photoplate} depicts one of the early spectral photoplates from the Hamburg Observatory, recorded in 1920. Pickering was very committed to that project and in 1890 the \emph{Draper Catalogue of Stellar Spectra} was published, containing spectroscopic classifications of 10,351 stars by Williamina Fleming (1857 -- 1911).

In 1890 the HCO put into operation a new observatory in Arequipa, Peru, and thus stars of the southern hemisphere were accessible for the ongoing campaign, too. Annie Jump Cannon (1863 -- 1941) analyzed the plates from this observatory and in 1901 published her first catalog of spectra of bright southern stars as part of the Henry Draper Memorial. In the following years, Cannon worked on a new classification scheme, which became known as the \emph{Harvard Classification} and which she applied to the stars in the \emph{Henry Draper Catalog}, published between 1918 and 1924 \citep{hco1971}. Annie Jump Cannon analyzed stars from the spectral photoplates until she died in 1941. Counting the Henry Draper catalog and its two extensions, she manually classified more than 350,000 stars, more than anyone else. Her catalogs paved the way for the later Morgan--Keenan classification, which still is in use today and gives access to stellar properties like surface temperature, surface chemical composition, surface gravity, as well as radius and mass with the use of stellar modeling.

Combining time domain astronomy and stellar spectroscopy, i.e., analyzing time series of stellar spectra, gives further insight into variable stars.
In 1889 Pickering detected a changing double absorption line while observing Mizar A. Mizar A is a double star system, that can not visually be resolved. However, the spectral data reveal the binary character of this object. In the same year Hermann Carl Vogel (1841 -- 1907) discovered that Algol is also a spectroscopic binary star system --- the year 1889 marks the discovery of \emph{spectroscopic binary stars}.

\section{Modern Stellar Astronomy}\label{sec_modern_astronomy}

Two major improvements are used in modern observatories for stellar astronomy. First, the CCD and CMOS solid-state detectors have a linear characteristic and a quantum efficiency of more than 90 \%, allowing them to capture rapidly changing signals, e.g., fast transients, and to study very faint sources. Second, atmospheric distortion can be overcome to a large extent by applying active and adaptive optics.

The Vera C. Rubin Observatory will house the Simonyi Survey Telescope, one of the most advanced survey telescopes --- a very wide field of view and the largest digital camera so far, really a worthy successor of the 20th-century photographic plate surveys. It will image the entire visible sky from Cerro Pachón, Chile, every few nights for ten years, observe about 17 billion stars, and thus monitor their changes in brightness and positions. The Vera C. Rubin Observatory will deliver very precise high-cadence light curves of more variable stars, that we know so far. The full survey operations are expected to begin in August 2025.

A major advance in spectroscopy was already used in an early design by Fraunhofer: a diffraction grating spectrometer. Modern spectrometers are stabilized and use very accurate calibration sources, e.g., frequency combs, and reach resolving powers larger than 100,000 (e.g.\ HARPS, ESPRESSO). For most of the features interesting to stellar spectroscopy, a resolving power below 100,000 is sufficient. Multi-object spectroscopic telescopes use optical fibers to feed the light of many different sources into a slit spectrometer, used e.g.,\ by SDSS and LAMOST. With a spectral resolving power of 21,000 in the visible range, the 4MOST facility will be able to observe 2436 objects simultaneously. This telescope will start its operation in mid-2024 and will take spectra of thousands of stars within the Milky Way and its neighboring satellite galaxies. 

Space telescopes further expand the observation prospects. First, they allow continuous observations without interference from solar irradiation and atmospheric distortions. The HIPPARCOS satellite in the 1990s was an astrometry mission that reduced the errors in distances to stars by about one order of magnitude. Its successor, the Gaia satellite \citep{Gaiamission} was launched in December 2013 and is still in operation. Up today, it mapped 1.8 billion stars of the Milky Way and achieved an accuracy of the parallax in the range of 20 micro arc seconds and will further improve the data in the next releases \citep{hoeg2024}. Three-dimensional mapping of the stars within our galaxy and their precise characterization will not only produce an enhanced vision of stellar evolution but also exploit the influence of the galactic environment on stellar evolution and vice versa.

And, space telescopes open the window to the higher energy spectral range. Especially the final stages of stellar evolution of more massive stars produce high energy radiation in the X-ray and gamma-ray range and can be studied with space telescopes, e.g. XMM-Newton, Chandra, Swift, INTEGRAL, and Fermi Gamma-ray Space Telescope.

Ground-based observatories using the near-infrared and sub-mm spectral range are used to study the formation of stars. The most productive recent sub-mm observatory is ALMA. High energy gamma-ray emission e.g., from pulsars, are very successfully observed by atmospheric Cherenkov telescopes, e.g., HESS, Veritas, and MAGIC. 

Last but not least in 2016 the first successful detection of a gravitational wave from a black hole merger was reported. For stellar astrophysics more important are mergers of neutron stars, so-called kilonova events. In 2017 the Fermi satellite observed a gamma-ray burst, while 6 min later a gravitational wave candidate was detected that showed typical properties of a binary neutron star merger. An extensive campaign was started to check for optical transients, which finally was discovered less than 11 hours after the merger. This was one of the first examples of a \emph{multi--messenger observation}. The combined interpretation of signals from different types of observations will lead to a much better understanding of these high-energy events in stellar life.

\section{Conclusion}\label{sec_conclusion}

There are still many open questions, many unsolved puzzles about stars, e.g., the distribution of stars of different masses upon creation in a molecular cloud --- the so-called \emph{initial mass function}, birth, life, and ending of massive stars, creation of the distribution of elements during the life cycle of stars, etc. The new Vera C. Rubin Observatory will deliver high-cadence data and create many alerts for follow-up observations. These and the large number of stellar spectra from 4MOST among others are expected to help to get deeper insights into the processes within the stars of our Universe.

\begin{ack}[Acknowledgments]
	
I thank Dr. Rob H. van Gent, Utrecht University, for his critical and very helpful comments.
\end{ack}


\begin{thebibliography*}{32}
	\providecommand{\bibtype}[1]{}
	\providecommand{\natexlab}[1]{#1}
	{\catcode`\|=0\catcode`\#=12\catcode`\@=11\catcode`\\=12
		|immediate|write|@auxout{\expandafter\ifx\csname
			natexlab\endcsname\relax\gdef\natexlab#1{#1}\fi}}
	\renewcommand{\url}[1]{{\tt #1}}
	\providecommand{\urlprefix}{URL }
	\expandafter\ifx\csname urlstyle\endcsname\relax
	\providecommand{\doi}[1]{doi:\discretionary{}{}{}#1}\else
	\providecommand{\doi}{doi:\discretionary{}{}{}\begingroup
		\urlstyle{rm}\Url}\fi
	\providecommand{\bibinfo}[2]{#2}
	\providecommand{\eprint}[2][]{\url{#2}}
	
	\bibtype{Book}%
	\bibitem[Aitken(1918)]{aitken1918}
	\bibinfo{author}{Aitken RG} (\bibinfo{year}{1918}).
	\bibinfo{title}{The Binary Stars}, \bibinfo{publisher}{McGraw-Hill Book
		Company}, \bibinfo{address}{New York}.
	
	\bibtype{incollection}%
	\bibitem[Argelander(1844)]{argelaner1844}
	\bibinfo{author}{Argelander FWA} (\bibinfo{year}{1844}),
	\bibinfo{title}{Aufforderung an {F}reunde der {A}stronomie, zur {A}nstellung
		von eben so interessanten und nützlichen, als leicht auszuführenden
		{B}eobachtungen über mehrere wichtige {Z}weige der {H}immelskunde},
	\bibinfo{editor}{Schumacher H}, (Ed.), \bibinfo{booktitle}{Schumacher's
		Jahrbuch für 1844}, \bibinfo{publisher}{J.G. Cotta},
	\bibinfo{address}{Stuttgart \& Tübingen}, pp. \bibinfo{pages}{122}.
	
	\bibtype{Book}%
	\bibitem[Bessel(1830)]{bessel1830}
	\bibinfo{author}{Bessel FW} (\bibinfo{year}{1830}).
	\bibinfo{title}{Tabulae Regiomontanae reductionum observationum astronomicarum
		ab anno 1750 AD 1850}, \bibinfo{publisher}{Regiomonti Prussorum},
	\bibinfo{address}{Paris, London}.
	
	\bibtype{Article}%
	\bibitem[Bessel(1839)]{bessel1839}
	\bibinfo{author}{Bessel FW} (\bibinfo{year}{1839}).
	\bibinfo{title}{Bestimmung der {E}ntfernung des 61sten {S}terns des {S}chwans}.
	\bibinfo{journal}{{\em Astronomische Nachrichten}} \bibinfo{volume}{16}:
	\bibinfo{pages}{65}.
	
	\bibtype{Book}%
	\bibitem[Clerke(1902)]{clerke1902}
	\bibinfo{author}{Clerke AM} (\bibinfo{year}{1902}).
	\bibinfo{title}{A Popular History of Astronomy during the Nineteenth Century},
	\bibinfo{edition}{fourth} ed., \bibinfo{publisher}{Adam and Charles Black},
	\bibinfo{address}{London}.
	
	\bibtype{Article}%
	\bibitem[Cunningham(2020)]{cunningham2020}
	\bibinfo{author}{Cunningham CJ} (\bibinfo{year}{2020}).
	\bibinfo{title}{'{D}ark stars' and a new interpretation of the ancient {G}reek
		stellar magnitude system}.
	\bibinfo{journal}{{\em Journal of Astronomical History and Heritage}}
	\bibinfo{volume}{23}: \bibinfo{pages}{231}.
	
	\bibtype{Book}%
	\bibitem[Eichhorn(1974)]{eichhorn1974}
	\bibinfo{author}{Eichhorn HK} (\bibinfo{year}{1974}).
	\bibinfo{title}{Astronomy of star positions: A critical investigation of star
		catalogues, the methods of their construction, and their purposes},
	\bibinfo{publisher}{Ungar}, \bibinfo{address}{New York}.
	
	\bibtype{Article}%
	\bibitem[Enke et al.(2024)]{Enke2024}
	\bibinfo{author}{Enke H}, \bibinfo{author}{Tuvikene T}, \bibinfo{author}{Groote
		D}, \bibinfo{author}{Edelmann H} and  \bibinfo{author}{Heber U}
	(\bibinfo{year}{2024}).
	\bibinfo{title}{Archives of photographic plates for astronomical use
		(applause). digitisation of astronomical plates and their integration into
		the international virtual observatory}.
	\bibinfo{journal}{{\em Astronomy \& Astrophysics}}
	\bibinfo{volume}{687}:
	\bibinfo{pages}{A165}.
	
	\bibtype{Article}%
	\bibitem[Fricke(1985)]{fricke1985}
	\bibinfo{author}{Fricke W} (\bibinfo{year}{1985}).
	\bibinfo{title}{Fundamental {C}atalogues -- {P}ast {P}resent and {F}uture}.
	\bibinfo{journal}{{\em Celestial Mechanics}} \bibinfo{volume}{36}:
	\bibinfo{pages}{207}.
	
	\bibtype{Article}%
	\bibitem[{Gaia Collaboration}(2016)]{Gaiamission}
	\bibinfo{author}{{Gaia Collaboration}} (\bibinfo{year}{2016}).
	\bibinfo{title}{The {G}aia mission}.
	\bibinfo{journal}{{\em Astronomy \& Astrophysics}} \bibinfo{volume}{595}:
	\bibinfo{pages}{A1}.
	
	\bibtype{Article}%
	\bibitem[Gysembergh et al.(2022)]{Gysembergh2022}
	\bibinfo{author}{Gysembergh V}, \bibinfo{author}{Williams PJ} and
	\bibinfo{author}{Zingg E} (\bibinfo{year}{2022}).
	\bibinfo{title}{New evidence for {H}ipparchus’ {S}tar {C}atalogue revealed by
		multispectral imaging}.
	\bibinfo{journal}{{\em Journal for the History of Astronomy}}
	\bibinfo{volume}{53}: \bibinfo{pages}{383--393}.
	
	\bibtype{Misc}%
	\bibitem[{Hoffleit} and {Warren}(1995)]{Hoffleit1991}
	\bibinfo{author}{{Hoffleit} D} and  \bibinfo{author}{{Warren} W.~H. J}
	(\bibinfo{year}{1995}), \bibinfo{month}{Nov.}
	\bibinfo{title}{{VizieR Online Data Catalog: Bright Star Catalogue, 5th Revised
			Ed.}}
	\bibinfo{howpublished}{VizieR On-line Data Catalog: V/50. Originally published
		in: 1964BS....C......0H; 1991bsc..book.....H}.
	
	\bibtype{Book}%
	\bibitem[Hoffmann(2017)]{Hoffmann2017}
	\bibinfo{author}{Hoffmann SM} (\bibinfo{year}{2017}).
	\bibinfo{title}{Hipparchs {H}immelsglobus}, \bibinfo{publisher}{Springer
		Fachmedien}, \bibinfo{address}{Wiesbaden}.
	
	\bibtype{Article}%
	\bibitem[H{\o}g(2017)]{hoeg2017}
	\bibinfo{author}{H{\o}g E} (\bibinfo{year}{2017}).
	\bibinfo{title}{Selected astrometric catalogues}.
	\bibinfo{journal}{{\em arXiv e-prints}}
	\bibinfo{doi}{\doi{10.48550/arXiv.2405.02017}}.
	\bibinfo{url}{\url{https://arxiv.org/abs/1706.08097}}.
	
	\bibtype{Article}%
	\bibitem[H{\o}g(2024)]{hoeg2024}
	\bibinfo{author}{H{\o}g E} (\bibinfo{year}{2024}).
	\bibinfo{title}{A review of 70 years with astrometry}.
	\bibinfo{journal}{{\em Astrophysics and Space Science}} \bibinfo{volume}{369}:
	\bibinfo{pages}{23}.
	
	\bibtype{Book}%
	\bibitem[Horrebow(1735)]{horrebow1735}
	\bibinfo{author}{Horrebow P} (\bibinfo{year}{1735}).
	\bibinfo{title}{Basis astronomiae sive astronomiae pars mechanica in qua
		Danica; simulque eorundem usus, sive methodi observandi Roemerianae
		describuntur observatoria, atque instrumenta astronomica Roemeriana},
	\bibinfo{publisher}{Christiani Paulli}, \bibinfo{address}{Kopenhagen}.
	
	\bibtype{Book}%
	\bibitem[Jones and Boyd(1971)]{hco1971}
	\bibinfo{author}{Jones BZ} and  \bibinfo{author}{Boyd LG}
	(\bibinfo{year}{1971}).
	\bibinfo{title}{The Harvard College Observatory -- The First Four
		Directorships}, \bibinfo{publisher}{Harvard University Press},
	\bibinfo{address}{Harvard}.
	
	\bibtype{Article}%
	\bibitem[Knobel(1877)]{knobel1877}
	\bibinfo{author}{Knobel EB} (\bibinfo{year}{1877}).
	\bibinfo{title}{The {C}hronology of {S}tar {C}atalogues}.
	\bibinfo{journal}{{\em Memoirs of the Royal Astronomical Society}}
	\bibinfo{volume}{43}: \bibinfo{pages}{1}.
	
	\bibtype{incollection}%
	\bibitem[Kroll(2009)]{kroll2009}
	\bibinfo{author}{Kroll P} (\bibinfo{year}{2009}), \bibinfo{title}{Real and
		virtual heritage – the plate archive in sonneberg – digitisation,
		preservation and scientiﬁc programme}, \bibinfo{editor}{Wolfschmidt G},
	(Ed.), \bibinfo{booktitle}{Cultural Heritage of Astronomical Observatories:
		From Classical Astronomy to Modern Astrophysics}, \bibinfo{publisher}{hendrik
		Bäßler verlag}, \bibinfo{address}{Berlin}, pp. \bibinfo{pages}{311}.
	
	\bibtype{Article}%
	\bibitem[Lequeux(2014)]{lequex2014}
	\bibinfo{author}{Lequeux J} (\bibinfo{year}{2014}).
	\bibinfo{title}{From {F}lamsteed to {P}iazzi and {L}alande: new standards in
		18th century astrometry}.
	\bibinfo{journal}{{\em Astronomy \& Astrophysics}} \bibinfo{volume}{567}:
	\bibinfo{pages}{A26}.
	
	\bibtype{Book}%
	\bibitem[Linton(2004)]{linton2004}
	\bibinfo{author}{Linton CM} (\bibinfo{year}{2004}).
	\bibinfo{title}{From Eudoxus to Einstein -- A History of Mathematical
		Astronomy}, \bibinfo{publisher}{Cambridge University Press},
	\bibinfo{address}{Cambridge}.
	
	\bibtype{Article}%
	\bibitem[Michell(1767)]{michell1767}
	\bibinfo{author}{Michell J} (\bibinfo{year}{1767}).
	\bibinfo{title}{An {I}nquiry into the probable {P}arallax, and {M}agnitude of
		the fixed {S}tars, from the {Q}uantity of {L}ight which they afford us, and
		the particular {C}ircumstances of their {S}ituation}.
	\bibinfo{journal}{{\em Philosophical Transactions}} \bibinfo{volume}{57}:
	\bibinfo{pages}{234}.
	
	\bibtype{Article}%
	\bibitem[Perryman(2012)]{perryman2012}
	\bibinfo{author}{Perryman M} (\bibinfo{year}{2012}).
	\bibinfo{title}{The history of astrometry}.
	\bibinfo{journal}{{\em Eur. Phys. J. H}} \bibinfo{volume}{37}:
	\bibinfo{pages}{745}.
	
	\bibtype{Article}%
	\bibitem[Pogson(1856)]{pogson1856}
	\bibinfo{author}{Pogson N} (\bibinfo{year}{1856}).
	\bibinfo{title}{Magnitudes of {T}hirty-six of the {M}inor {P}lanets for the
		{F}irst {D}ay of each {M}onth of the {Y}ear 1857}.
	\bibinfo{journal}{{\em Monthly Notices of the Royal Astronomical Society}}
	\bibinfo{volume}{17}: \bibinfo{pages}{12}.
	
	\bibtype{Article}%
	\bibitem[Protte and Hoffmann(2020)]{protte2020}
	\bibinfo{author}{Protte P} and  \bibinfo{author}{Hoffmann SM}
	(\bibinfo{year}{2020}).
	\bibinfo{title}{Accuracy of magnitudes in pre-telescopic star catalogs}.
	\bibinfo{journal}{{\em Astronomische Nachrichten}} \bibinfo{volume}{341}:
	\bibinfo{pages}{827}.
	
	\bibtype{Book}%
	\bibitem[Ruggles(2015)]{ruggles2015}
	\bibinfo{editor}{Ruggles CLN}, (Ed.)  (\bibinfo{year}{2015}).
	\bibinfo{title}{Handbook of Archaeoastronomy and Ethnoastronomy},
	\bibinfo{publisher}{Springer}, \bibinfo{address}{New York}.
	
	\bibtype{Inproceedings}%
	\bibitem[Schrimpf and Verbunt(2021)]{schrimpf2021}
	\bibinfo{author}{Schrimpf A} and  \bibinfo{author}{Verbunt F}
	(\bibinfo{year}{2021}), \bibinfo{title}{The star catalogue of {W}ilhelm {IV},
		{L}andgraf von {H}essen-{K}assel}, \bibinfo{editor}{Wolfschmidt G} and
	\bibinfo{editor}{Hoffmann SM}, (Eds.), \bibinfo{booktitle}{Applied and
		{C}omputational {H}istorical {A}stronomy -- Proceedings of the Splinter
		Meeting of the Astronomische Gesellschaft}, \bibinfo{volume}{55},
	\bibinfo{publisher}{tredition GmbH}, \bibinfo{address}{Hamburg}, pp.
	\bibinfo{pages}{176}, \bibinfo{url}{\url{https://arxiv.org/abs/2103.10801}}.
	
	\bibtype{Article}%
	\bibitem[Verbunt and Schrimpf(2021)]{verbunt2021}
	\bibinfo{author}{Verbunt F} and  \bibinfo{author}{Schrimpf A}
	(\bibinfo{year}{2021}).
	\bibinfo{title}{The star catalogue of {W}ilhelm {IV}, {L}andgraf von
		{H}essen-{K}assel --- {A}ccuracy of the catalogue and of the measurements}.
	\bibinfo{journal}{{\em Astronomy \& Astrophysics}} \bibinfo{volume}{649}:
	\bibinfo{pages}{A112}.
	
	\bibtype{Article}%
	\bibitem[Verbunt and van~der Sluys(2019)]{verbunt2019}
	\bibinfo{author}{Verbunt F} and  \bibinfo{author}{van~der Sluys M}
	(\bibinfo{year}{2019}).
	\bibinfo{title}{Why {H}alley {D}id {N}ot {D}iscover {P}roper {M}otion and {W}hy
		{C}assini {D}id}.
	\bibinfo{journal}{{\em Journal for the History of Astronomy}}
	\bibinfo{volume}{50}: \bibinfo{pages}{383}.
	
	\bibtype{Article}%
	\bibitem[Verbunt and van Gent(2010{\natexlab{a}})]{verbunt2010hevelius}
	\bibinfo{author}{Verbunt F} and  \bibinfo{author}{van Gent RH}
	(\bibinfo{year}{2010}{\natexlab{a}}).
	\bibinfo{title}{The star catalogue of {H}evelius --- {M}achine-readable version
		and comparison with the modern {H}ipparcos {C}atalogue}.
	\bibinfo{journal}{{\em Astronomy \& Astrophysics}} \bibinfo{volume}{516}:
	\bibinfo{pages}{A29}.
	
	\bibtype{Article}%
	\bibitem[Verbunt and van Gent(2010{\natexlab{b}})]{verbunt2010}
	\bibinfo{author}{Verbunt F} and  \bibinfo{author}{van Gent RH}
	(\bibinfo{year}{2010}{\natexlab{b}}).
	\bibinfo{title}{Three editions of the star catalogue of {T}ycho {B}rahe ---
		{M}achine-readable versions and comparison with the modern {H}ipparcos
		{C}atalogue}.
	\bibinfo{journal}{{\em Astronomy \& Astrophysics}} \bibinfo{volume}{516}:
	\bibinfo{pages}{A28}.
	
	\bibtype{Article}%
	\bibitem[Verbunt and van Gent(2012)]{verbunt2012}
	\bibinfo{author}{Verbunt F} and  \bibinfo{author}{van Gent RH}
	(\bibinfo{year}{2012}).
	\bibinfo{title}{The star catalogues of {P}tolemaios and {U}lugh {B}eg ---
		{M}achine-readable versions and comparison with the modern {HIPPARCOS}
		{C}atalogue}.
	\bibinfo{journal}{{\em Astronomy \& Astrophysics}} \bibinfo{volume}{544}:
	\bibinfo{pages}{A31}.
	
\end{thebibliography*}

\end{document}